\documentclass[unsortedaddress,
onecolumn,preprintnumbers,amsmath,amssymb]{revtex4}
\usepackage{graphicx}% Include Figure files     fleqn,
\usepackage{dcolumn}% Align table columns on decimal point
\usepackage{bm}% bold math
\usepackage{latexsym}

\begin{document}

%\begin{center}
%\Large
\title{Aspects of electron transport through a quantum dot}

%\vspace{0.5cm}
\large
\author{ R. Taranko, M. Krawiec, T. Kwapi\'{n}ski, E. Taranko,\\
        K.~I.~Wysoki\'{n}ski \\
Institute of Physics and Nanotechnology Center, \\
Maria
Curie-Sk\l odowska University,\\
Pl. Marii Curie-Sk\l odowskiej 1,
20-031 Lublin, Poland}

%\end{center}

\date{\today}
%\maketitle
%%%%%%%%%%%%%%%%%%%%%%%%%%%%%%%%%%%%%%%%%%%%%%%%%

\vspace{0.5cm}
\begin{abstract}
The work of the Lublin group   on the non-equillibrium transport
 through the quantum dot coupled to external leads
 (normal or superconducting) and subject to external time
 dependent fields has been reviewed.
\end{abstract}

%%%%%%%%%%%%%%%%%%%%%%%%%%%%%%%%%%%%%%%%%%%%%%%%%%%%%%%%%%%%%%%%%%%%%%%%%%%%%%%
\maketitle

\vspace{1cm}
\section{Introduction}
This paper is an extended version of the talk given at
 the LFPPI-network conference in Lodz (Poland).
It makes extensive use of our published work
\cite{Lublin_qd1}-\cite{Lublin_qd9}.
 In particular all figures which appear in the present review
 have already been presented there.

Due to recent developments in fabrication of small electronic devices
equilibrium and non-equilibrium properties of mesoscopic systems have been of
great interest during last decade. In particular a lot
of work has been done on electron transport through
quantum dots.

The quantum dots are small pieces of matter containing few to few tens of
electrons. Very often they are electrically defined
in a two dimensional electron
gas placed at the interface in GaAs-GaAlAs heterostructure. The
important point about quantum dots is that they can be viewed like
impurities in solids \cite{PALee} but at the same time
their coupling to external
environment can be controlled and the noneguilibrium
transport\cite{non-eq} can be studied.

The discovery of the Kondo effect in quantum dots
\cite{GoldhaberGordon,Cronenwett} connected to external leads by tunnel
junctions has resulted in increased experimental and
theoretical interest. The
Kondo effect in quantum dots manifests itself
as an increased conductance $G$ of
the system at temperatures lower than
the Kondo temperature $T_K$. It is due to
formation of the so-called Abrikosov-Suhl
or Kondo resonance at the Fermi energy.
This is a many-body singlet state involving spin on the quantum dot and
electrons in the leads.

Here we review our studies of stationary nonequilibrium  transport with
help of Keldysh Green's function formalism. To obtain the
information about the
time dependent phenomena in a quantum dots which
parameters  do depend on time we calculate the time evolution
operator and then time dependent charge and current.

%%%%%%%%%%%%%%%%%%%%%%%%%%%%%%%%%%%%%%%%%%%%%%%%%%

\section{Time-independent transport}

In this section we shall present the results for time independent
 transport through quantum dot
with a single energy level. We start with asymmetrically coupled
quantum dot (section II.A),   discuss
 the role of Van Hove
singularity in the density of states of the electrodes (section
II.B) and consider the dot coupled to one normal and other
superconducting lead (section II.C).
\begin{figure}[t,b,h]
\begin{center}
 \resizebox{0.5\columnwidth}{!}{
  \includegraphics{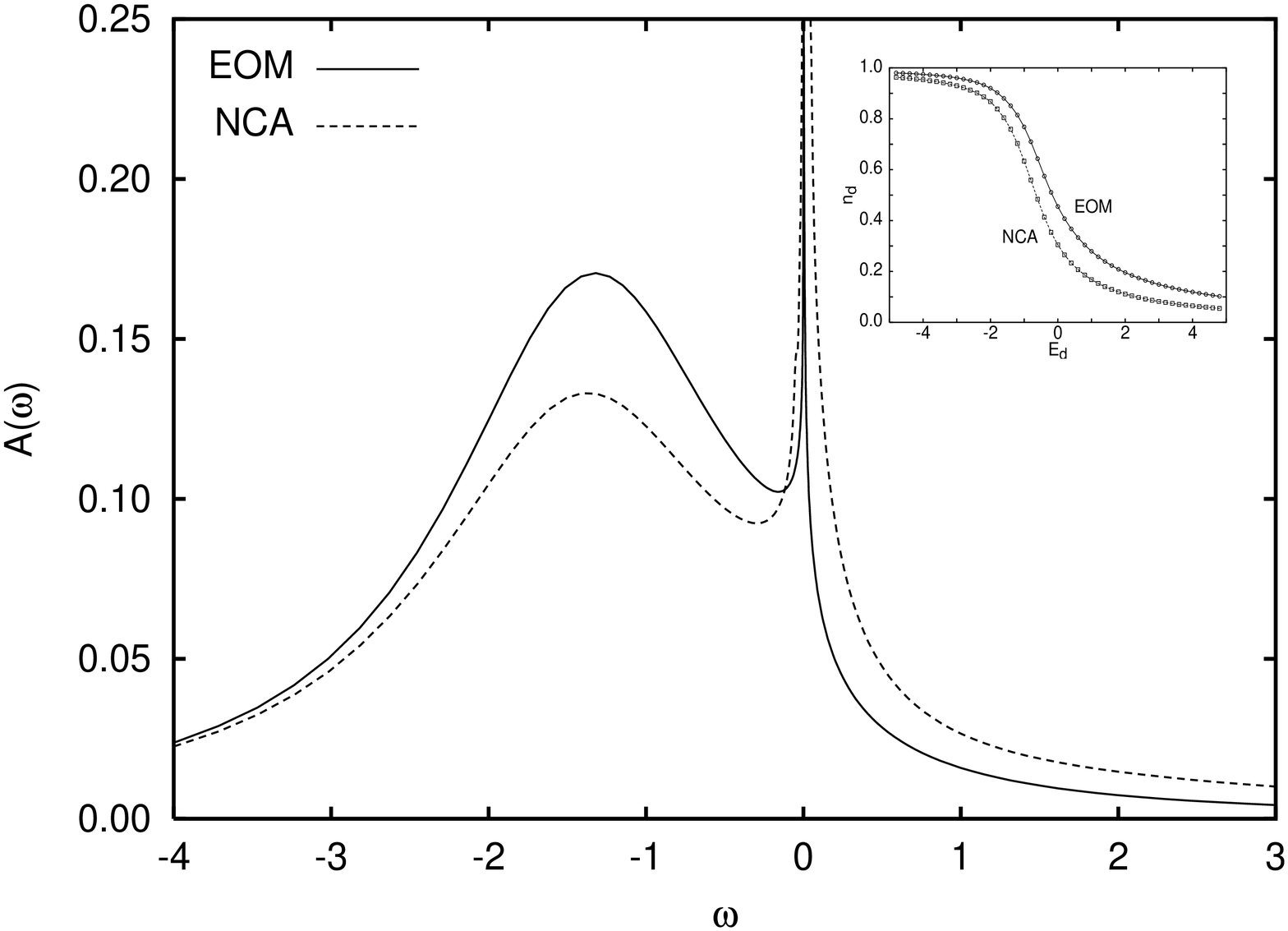}}
\end{center}
\caption{The equilibrium density of states on the quantum dot
          obtained within $EOM$ (solid line) and $NCA$ (dashed line). Note the
          relative shift of the spectral weight with respect to the chemical
          potential $\mu=0$ which results in different occupations shown in the
          inset. The parameters are: $E_d=-2$, $\Gamma_L = \Gamma_R = 1$ and
          $T=10^{-3}$. \cite{MKKIW_PRB}}
\label{Fig1}
\end{figure}

\subsection{Non-equilibrium Kondo effect
in asymmetrically coupled quantum dots}

It is well known that in systems containing quantum dot the Kondo effect
manifests itself (at low temperatures) as an enhancement of the conductance at
zero source-drain voltage\cite{PALee}, $V_{SD} = 0$. Occasionally the Kondo peak in
conductance appearing at non-zero voltages $V_{SD} \neq 0$ has been observed
\cite{Schmid,Simmel}, the so called anomalous Kondo effect. This phenomenon has
been investigated experimentally for the dot coupled weakly to one and strongly
to the other lead. It was observed that the source-drain voltage, at which the
peak appeared, scales roughly linearly with a gate voltage $V_g$. Performing
model calculations based on non-equilibrium transport theory we have found that
the emergence of the Kondo peak at non-zero voltages is caused by asymmetric
coupling to external leads.
\begin{figure}[h]
 \resizebox{0.4\columnwidth}{!}{
  \includegraphics{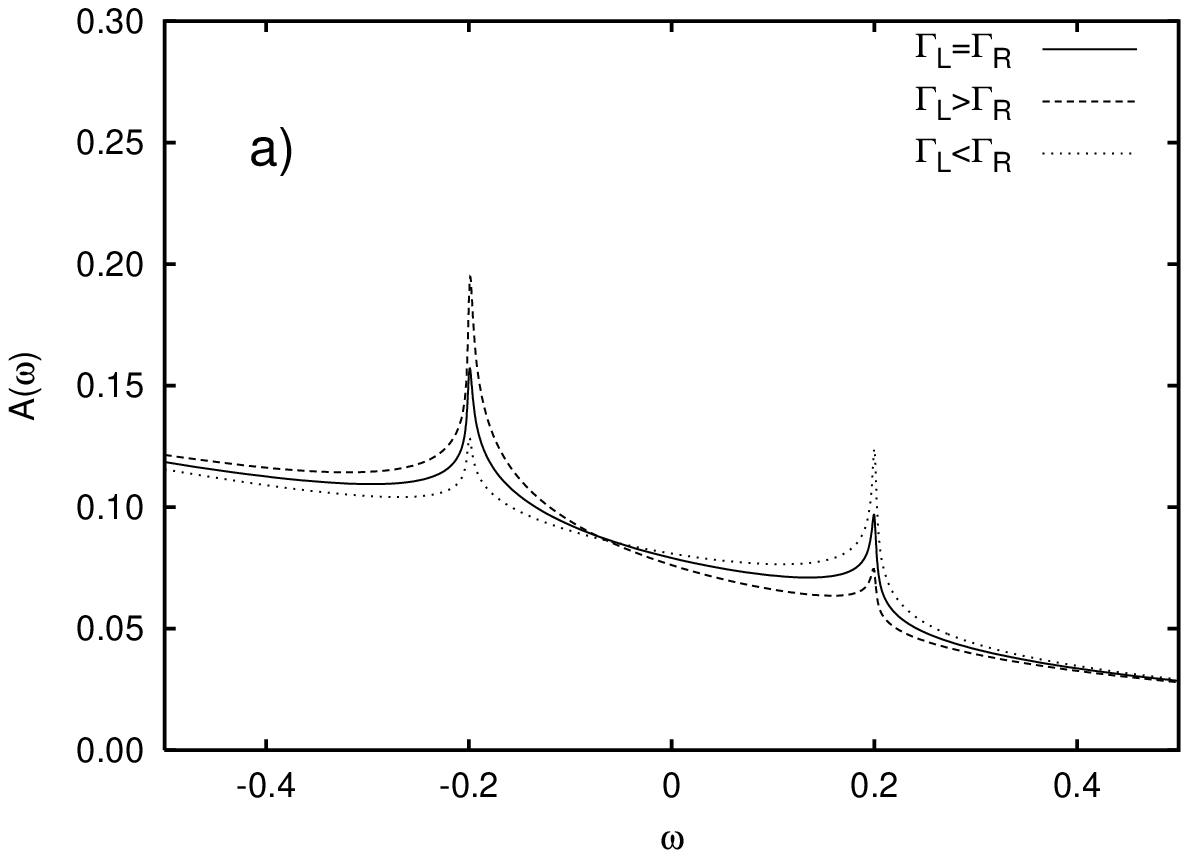}}
 \resizebox{0.4\columnwidth}{!}{
  \includegraphics{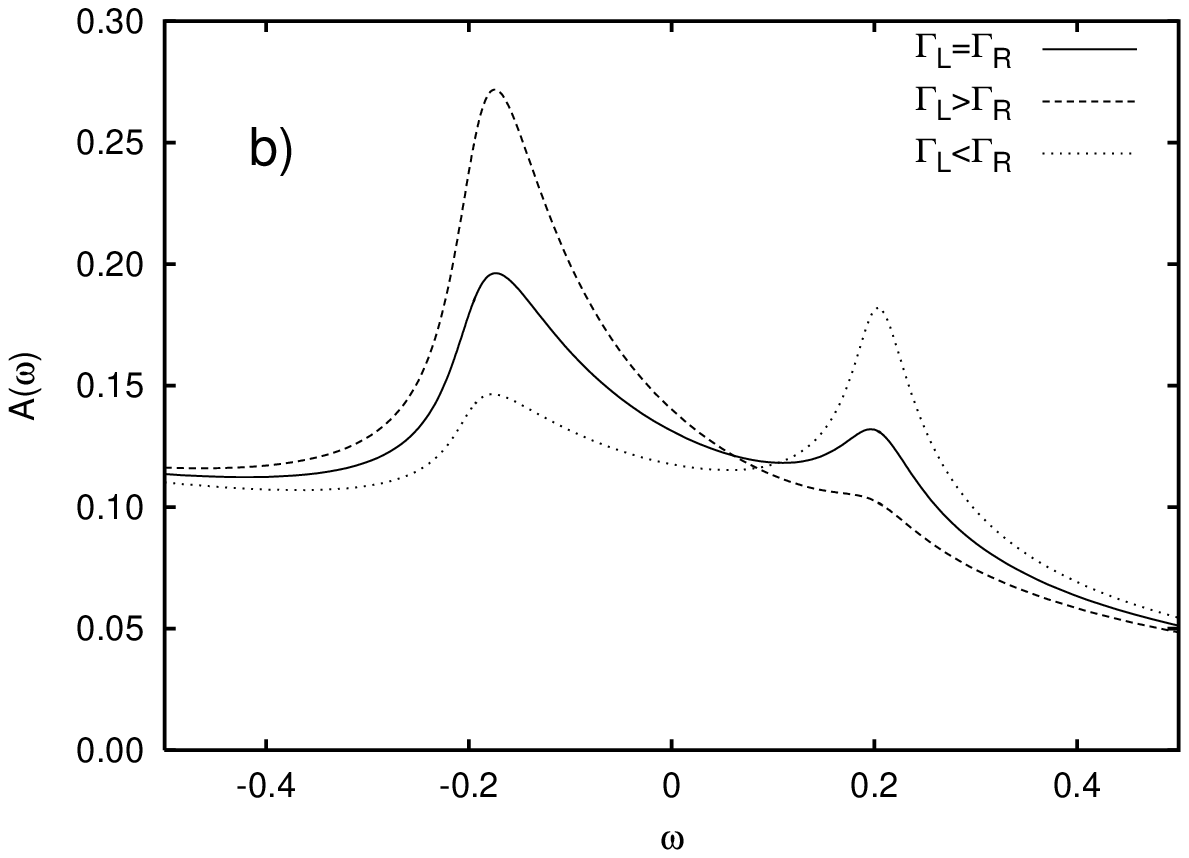}}
 \caption{\label{Fig2} The non-equilibrium density of states obtained within
          $a)$ - $EOM$ and $b)$ - $NCA$ for the symmetric
          $\Gamma_{L} = \Gamma_{R}$ (solid lines) and asymmetrically coupled
          quantum dot with $\Gamma_{L} = 2 \Gamma_{R}$ (dashed) and
          $\Gamma_L = \frac{1}{2} \Gamma_R$ (dotted lines).
          $\mu_R = - \mu_L = 0.2$ and the other parameters are the same as in
          Fig. (\ref{Fig1}). \cite{MKKIW_PRB}}
\end{figure}

The system under consideration is described by single impurity Anderson model
and the Landauer-type formula for the current flowing in the system,
left lead - quantum dot - right lead is \cite{non-eq}
\begin{eqnarray}
J = \frac{e}{\hbar} \sum_{\sigma} \int d\varepsilon
\frac{\Gamma^{\sigma}_L(\varepsilon) \Gamma^{\sigma}_R(\varepsilon)}
{\Gamma^{\sigma}_L(\varepsilon) + \Gamma^{\sigma}_R(\varepsilon)}
(- \frac{1}{\pi}) {\rm Im} G^r_{\sigma}(\varepsilon) [f_L(\varepsilon) -
f_R(\varepsilon)]
\label{Landauer}
\end{eqnarray}
where $f_{L/R}(\varepsilon)$ denotes the Fermi distribution function,
$G^r_{\sigma}(\varepsilon)$ - retarded on-dot Green's function and
$\Gamma^{\sigma}_{L/R}(\varepsilon)$ - effective coupling of the localized
electrons to the conduction band.

The retarded Green's function required for the calculations of the tunneling
current and the differential conductance of the system is obtained within the
equation of motion ($EOM$) technique with slave boson representation of the
electron operators and non-crossing approximation ($NCA$). In both cases we
assume $U \rightarrow \infty$. As we have checked, the $EOM$ method gives
correct position of the Kondo peak, however it leads to incorrect width of the
peak and occupation on the dot. Therefore we have used $NCA$ technique which is
generally accepted for description of the system in the Kondo regime.

In the Fig.\ref{Fig1} we present the equilibrium density of states ($DOS$) of
the quantum dot coupled to two leads obtained by means of the $NCA$ and $EOM$
approaches.
The main features of the $DOS$ remain the same
in both cases. However the height
and width of the Kondo peak are much larger in $NCA$.

Fig.\ref{Fig2} gives the non-equilibrium $DOS$ obtained for symmetric
($\Gamma_L = \Gamma_R$) and asymmetric couplings $\Gamma_L = 2 \Gamma_R$ and
$\Gamma_L = \frac{1}{2} \Gamma_R$.
We see that the Kondo peak is always located
at energies coinciding with those
of the Fermi levels of the leads \cite{non-eq}. Thus
in non-equilibrium we get two Kondo
resonances in the $DOS$ pinned to the Fermi energies
of corresponding leads.

\begin{figure}[h]
 \resizebox{0.4\columnwidth}{!}{
  \includegraphics{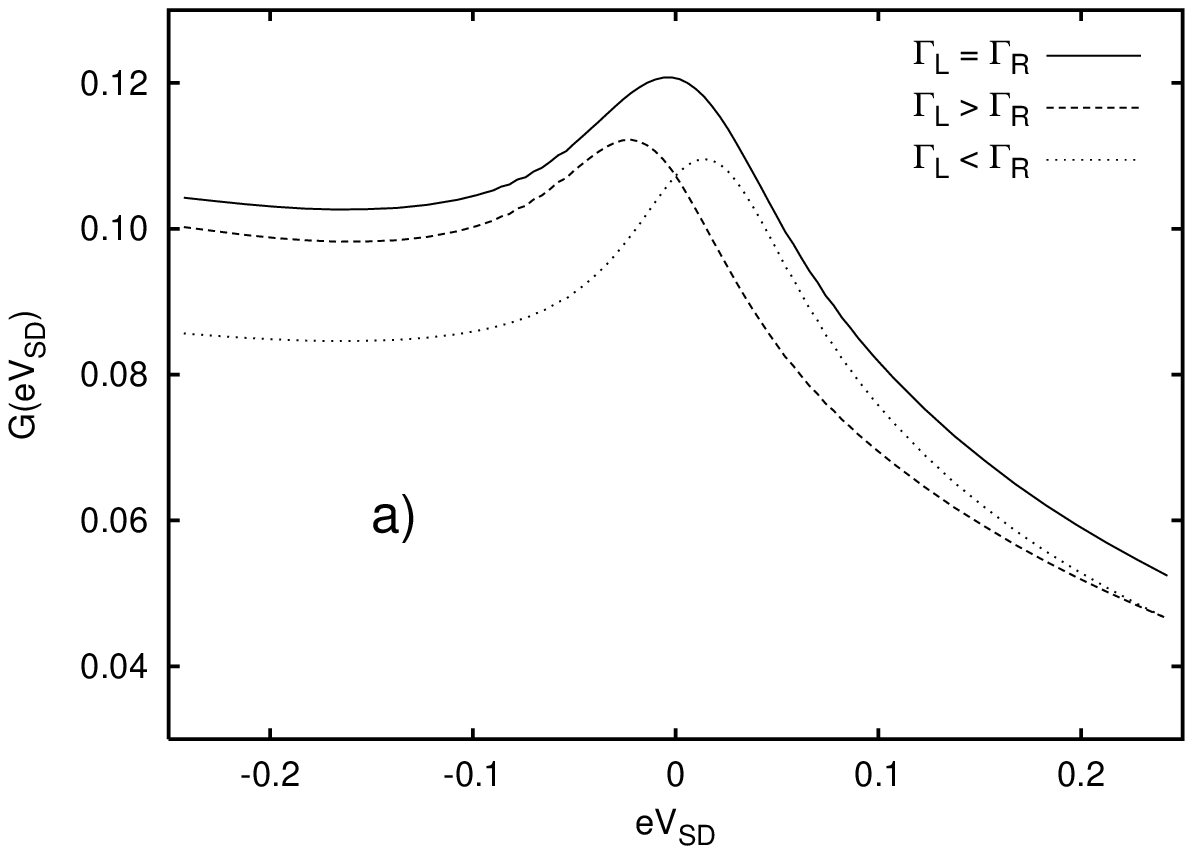}}
 \resizebox{0.4\columnwidth}{!}{
  \includegraphics{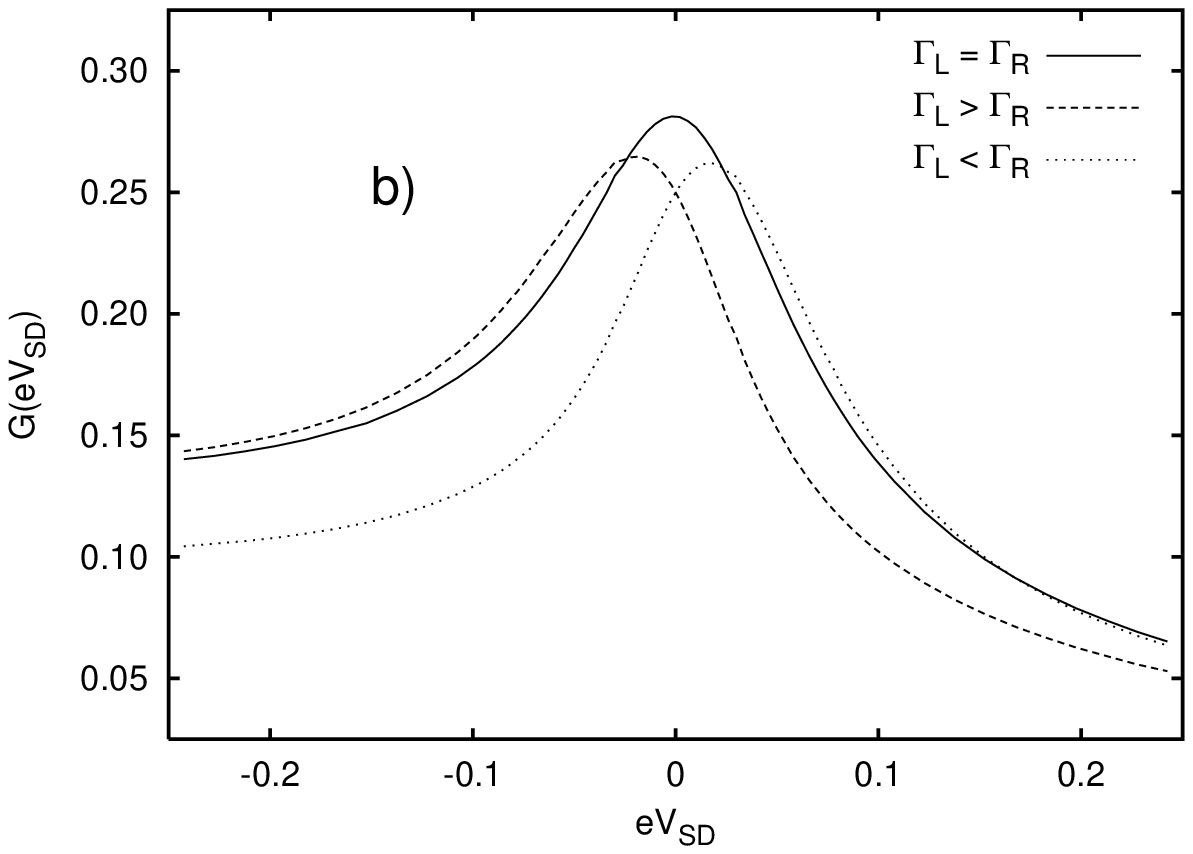}}
 \caption{\label{Fig3} The differential conductance
          ($G(eV_{SD}) = d J/d (eV_{SD}$)) obtained within
          $a)$ - $EOM$ and $b)$ - $NCA$ for the symmetric
          $\Gamma_{L} = \Gamma_{R}$ (solid lines) and asymmetrically coupled
          quantum dot with $\Gamma_{L} = 2 \Gamma_{R}$ (dashed) and
          $\Gamma_L = \frac{1}{2} \Gamma_R$ (dotted lines). \cite{MKKIW_PRB}}
\end{figure}

The differential conductance spectrum of the same system
is shown in the Fig.\ref{Fig3}.
For comparison we have also plotted the conductance of the symmetrically
coupled quantum dot. In the symmetric case ($\Gamma_L = \Gamma_R$) the
Kondo peak is located exactly at zero
bias ($V_{SD} = 0$) but for $\Gamma_L >
\Gamma_R$ ($\Gamma_L < \Gamma_R$) it is shifted to the negative (positive)
voltages. This finding is in nice qualitative agreement with experimental
situation in transport through the quantum
dot in the presence of asymmetric
barriers. Further details regarding
this anomalous Kondo effect can be found in \cite{MKKIW_PRB}.

\subsection{Van Hove singularities and the Kondo effect}

Here we present some results  regarding influence of the
band structure of the leads, in particular Van Hove
singularities in the $DOS$
on the tunneling current through the quantum dot. We have assumed the
conduction electron energies of the leads
in the form characteristic for two-dimensional
tight-binding spectrum which is known to possess the logarithmic
 singularity in
the middle of the band. To calculate the
retarded Green's function we have used
Hubbard I type decoupling and performed
calculations of the density of states
(DOS) on the dot and the differential conductunce trough it.
To see the
influence of spectrum of the electrode
on transport we repeated the calculations
for constant, Lorentzian
and Van Hove singular   $DOS$.

\begin{figure}[h]
\begin{center}
 \resizebox{0.5\columnwidth}{!}{
  \includegraphics{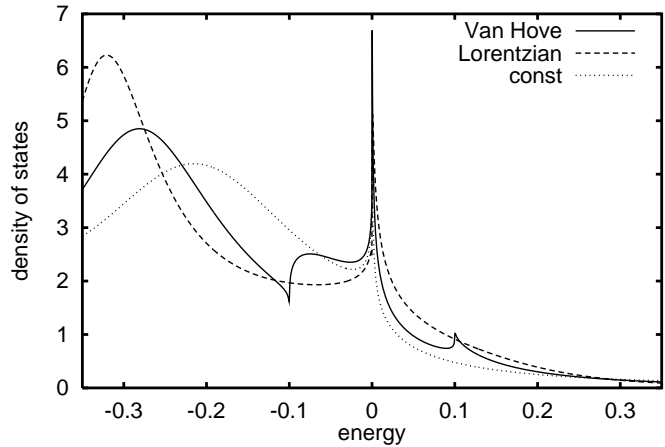}}
\end{center}
\caption{The density of states on the dot calculated for $V=0.1 W$,
$T=10^{-4} W$ ($W$ is the bandwidth) and for three different spectra of
electrons in the leads. Note the structure due to Van Hove singularities and
robustness of the coherent (Kondo) feature at the Fermi level. \cite{Lublin_qd7}}
\label{Fig4}
\end{figure}
\begin{figure}[h]
\begin{center}
 \resizebox{0.5\columnwidth}{!}{
  \includegraphics{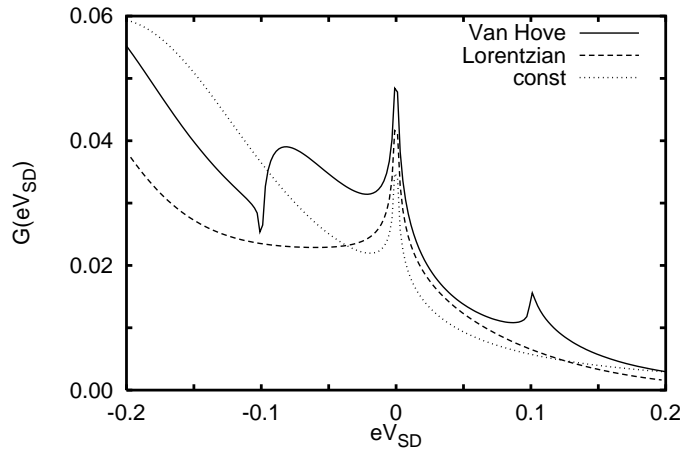}}
\end{center}
\caption{The differential conductance $G$ (in units of $e^2/h$) for three
spectra of electrons in the leads. The Van Hove singularity affects the
conductance and produces additional structure around the Fermi level. \cite{Lublin_qd7}}
\label{Fig5}
\end{figure}

Fig. \ref{Fig4}   shows the effect of the Van Hove singularity in the leads
on the density of states of the electrons on the dot
and figure \ref{Fig5} shows the corresponding
differential conductance spectra.
One observes marked changes of the tunneling current and the differential
conductance due to the Van Hove singularities (now shifted by $\pm 0.05 W$).
The dip seen below Fermi level ($\mu_L - \mu_R < 0$) is
due to the real part of
the self-energy, while the Kondo like structure above
the Fermi level (solid curve) is connected
with the imaginary part, which
is proportional to the lead $DOS$. The structure thus reflects the
presence of Van Hove singularity in
the spectrum of carriers in the leads.

\subsection{Electron transport through a quantum dot coupled to normal metal
and superconductor}

Here we present some of the results of our studies concerning transport
properties of the quantum dot coupled to the normal metal and $BCS$-like
superconductor ($N - QD - S$) in the presence of the Kondo effect and Andreev
scattering. The system is described by single impurity Anderson model in the
limit of strong on-dot interaction. We have used $EOM$ method to calculate
non-equilibrium Green's functions together with the modified slave boson
technique. The details of calculations can be found in
\cite{MKKIW_SST} and here we present the results for
the quantum dot $DOS$ only, as
it gives valuable information about the system.

\begin{figure}[h]
\begin{center}
 \resizebox{0.5\columnwidth}{!}{
  \includegraphics{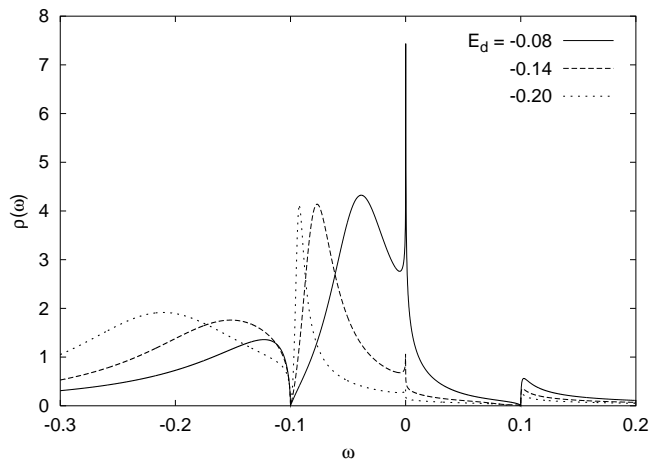}}
\end{center}
 \caption{\label{Fig6} The density of states of the quantum dot in
 $N-QD-S$ system for various
          values of the dot energy level $E_d$. Other parameters are
          following: $\Gamma_N  = \Gamma_S = 0.02$, $\Delta = 0.1$, $eV = 0$,
          $T = 10^{-5}$ in the units of the bandwidth $W$. \cite{MKKIW_SST}}
\end{figure}

In Fig. \ref{Fig6} we show the $DOS$ of
the quantum dot for various positions
of the dot energy level $E_d$.
Clearly, the Kondo effect manifests itself in the resonance
at the Fermi energy
which survives the presence of superconductivity in one electrode. The
additional structure at $\omega = \pm \Delta=\pm 0.1$ reflects
the superconducting gap $\Delta$.

\begin{figure}[h,b,t]
\begin{center}
 \resizebox{0.5\columnwidth}{!}{
  \includegraphics{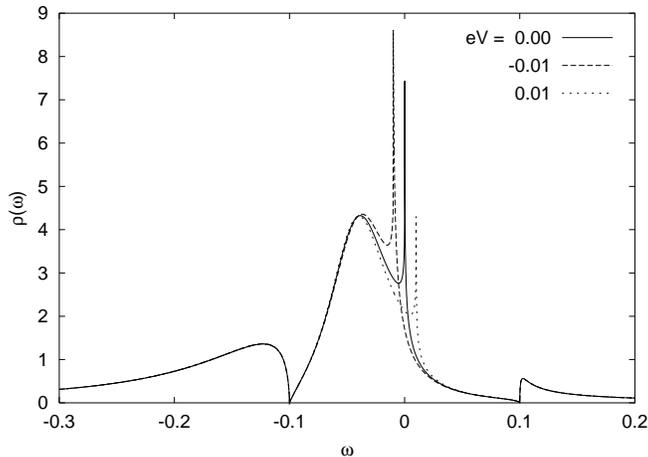}}
\end{center}
\caption{Equilibrium ($eV=0$) and non-equilibrium ($eV = \pm 0.01$)
 density of
states of the quantum dot in the $N-QD-S$ system.
Other parameters have following values:
$\Gamma_N = \Gamma_S = 0.02$, $\Delta = 0.1$,
$E_d = -0.08$, $T = 10^{-5}$ in
units of the bandwidth $W$. \cite{MKKIW_SST} }
\label{Fig7}
\end{figure}

\begin{figure}[t,h,b]
\begin{center}
 \resizebox{0.5\columnwidth}{!}{
  \includegraphics{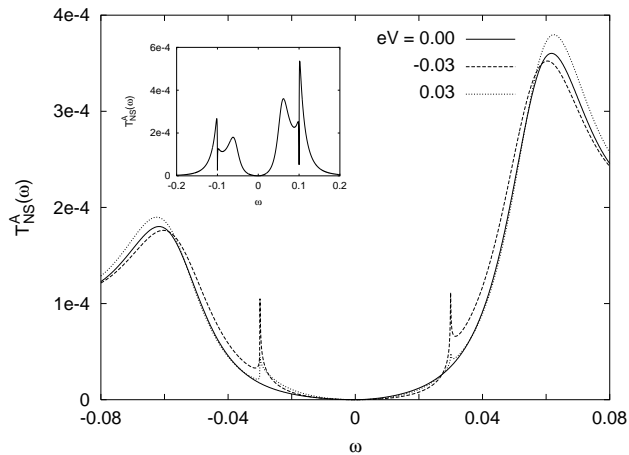}}
\end{center}
\caption{$T^A_{NS}(\omega)$ for different values of the bias voltage $eV=0$
(solid line), $-0.03$ (dashed) and $0.03$ (dotted line). $E_d = -0.08$,
$\Gamma_N = \Gamma_S = 0.01$ and
$\Delta = 0.1$. Inset: large scale view of the
equilibrium $T^A_{NS}(\omega)$ \cite{MKKIW_SST}.}
\label{Fig8}
\end{figure}

In the next step we consider $DOS$ in the non-equilibrium situation
($eV = \mu_N - \mu_S \neq 0$).
Recall that in the $N - QD - N$ system for  $eV \neq 0$
there emerge two resonances pinned
to Fermi levels of the left and right leads,
respectively. In the present case there is
a gap in the spectrum of the
superconducting lead and we observe only
one resonance pinned to the normal
metal Fermi level ($\mu_N$) (see Fig. \ref{Fig7}).
The presence of one superconducting electrode opens a possibility for
the Andreev reflections. This is process in which an electron
coming to the N-S interface from the normal metal side is reflected
back into metal as hole, while the Cooper pair enters the superconductor.
From the general expression for the nonequilibrium
current (\ref{Landauer}) in the $N-QD-S$ system one
expects the Kondo peak  in the tunneling mediated by the Andreev
reflections.   The
corresponding part of the total current
 is written in the form \cite{MKKIW_SST}
\begin{eqnarray}
J_A = - \frac{2 e}{h} \int^{\infty}_{-\infty} \frac{d\omega}{2\pi}
T^A_{NS}(\omega) [f(\omega - eV) - f(\omega + eV)],
\label{JAGNS}
\end{eqnarray}
where we have introduced ''transmittance''
$T^A_{NS}(\omega)$ associated with
the Andreev tunneling. At zero temperature and at energies less than the
superconducting gap, $T^A_{NS}(\omega)$ can be regarded as a total
transmittance because Andreev tunneling is
the only process allowed in these
circumstances. $T^A_{NS}(\omega)$ for different
values of the $eV$ is plotted
in the Fig.\ref{Fig8}.
The broad resonances at $\omega \approx \pm 0.06$
are reflections of the dot
energy level $E_d = 0.08$ for electrons and holes,
shifted from its original
position due to renormalization caused by the strong
Coulomb interaction.
The important point is that there is no Kondo peak
in equilibrium ($eV = 0$)
transmittance. However as soon as we go
away from the $eV = 0$, we can observe
the Kondo peaks at energies $\omega = \pm eV$ and the  strong
asymmetry between negative (dashed line) and positive
(dotted line) voltages.
While in the former case we have very well resolved
resonances, in the later
these resonances are strongly suppressed. This asymmetry
is strictly related to
the density of states (see Fig. (\ref{Fig6})), where we also observe such
asymmetry.

Since equilibrium transmittance $T^A_{NS}(\omega)$ does not
show the Kondo peak,
we cannot expect it in the differential conductance
$G_A(eV_{SD}) = d J_A/d (eV_{SD})$ with $J_A$ defined by (\ref{JAGNS}).
It turns out that $G_A(eV_{SD})$ is very
sensitive to the position of the dot
energy level (see Fig.7 in ref. \cite{MKKIW_SST}).
 More detailed discussion of these
 results can be found in Ref.
\cite{MKKIW_SST}.

%%%%%%%%%%%%%%%%%%%%%%%%%%%%%%%%%%%%%%%%%%%%%%%%%%%%%%%%%%%%%%%%%%%%%%
\section{\label{sec1}Time-dependent transport}
%%%%%%%%%%%%%%%%%%%%%%%%%%%%%%%%%%%%%%%%%%%%%%%%%%%%%%%%%%%%%%%%%%%%%%%

In this section we consider the transport properties of
a quantum dot (QD) under the influence of external
time-dependent fields. The high-frequency signals may be
applied to a QD and the time-dependent fields will modify
the tunneling current.
New effects have been observed and theoretically described,
e.g. photon-assisted tunneling through small quantum dots
with well-resolved discrete energy states \cite{1,2,3}, photon-electron
pumps \cite{4,5,6} and others.
One can investigate the current flowing through a QD under
periodic modulation of the QD electronic structure \cite{7} or
periodic (non-periodic) modulation of the tunneling barriers
\cite{6} and electron energy levels in both (left and right)
electron reservoirs \cite{8}

We consider the simplest case of the quantum dot
with multiple energy levels
without the electron-electron Coulomb interaction coupled to two leads
(right and left) in the presence of external microwave (MW) fields which
act on all parts of the system. Under the adiabatic approximation
the single-electron energies of the time dependent driven system can be
represented in the form $\varepsilon_{\vec k}(t) = \varepsilon_{\vec
k_\alpha} + \Delta\cos\omega t$.
We describe the dynamical evolution
of the charge localized on the QD and the current flowing through the system
in terms of the time evolution operator (for details see Refs [1,3,9,24].
%\cite{9}).

In the first subsection we consider the influence of the band structure of
the leads on electron transport through the QD and in the next subsection
we consider the time-dependent transport through the QD with additional
tunneling channel.

%%%%%%%%%%%%%%%%%%%%%%%%%%%%%%%%%%%%%%%%%%%%%%%%%%%%%%%%%%%%%%%%%%%%%%
\subsection{\label{subsec1}Band structure effects
in time-dependent electron transport
through the QD}
%%%%%%%%%%%%%%%%%%%%%%%%%%%%%%%%%%%%%%%%%%%%%%%%%%%%%%%%%%%%%%%%%%%%%%%

Here we are going to study the influence of the
singularities of the leads band structure
and therefore we are forced to go beyond WBL. Within the formalism of the
non-equilibrium Green's functions it is extremely difficult to calculate,
e.g. the time dependent current $j_L(t)$ tunneling from the left
 lead to the QD.
We recall that the general formula for the current flowing e.g. from
the left lead into the QD is usually expressed in terms of the QD
retarded Green function $G^r(t,t_1)$.
However, $G^r(t,t_1)$ satisfies the Dyson equation with subsequent double
time integrations and for the time-dependent
Hamiltonian it is a very difficult task to calculate it.
Therefore, in the following, we calculate the QD charge and current using
the evolution operator method.
The potential drop between the left and right leads is given by
$\mu_L - \mu_R = eV_{s-d}$ and $V_{s-d}$ is the measured voltage between
source and drain. In experiments the gate voltage
controls the position of QD's energy level $\epsilon_d$ and to mimic
measurements of the QD charge
or current vs gate voltage we have calculated them against $\epsilon_d$.
To study the effect of the leads band structure on the electron transport
 through the QD, we have assumed
the two-dimensional tight binding simple
cubic crystal spectrum (2D-TB) for leads conduction
electrons, which is known to possess the logarithmic
singularity in the middle of the band.

\begin{figure}[h,t,b]
\begin{center}
\resizebox{0.6\columnwidth}{!}{
\includegraphics{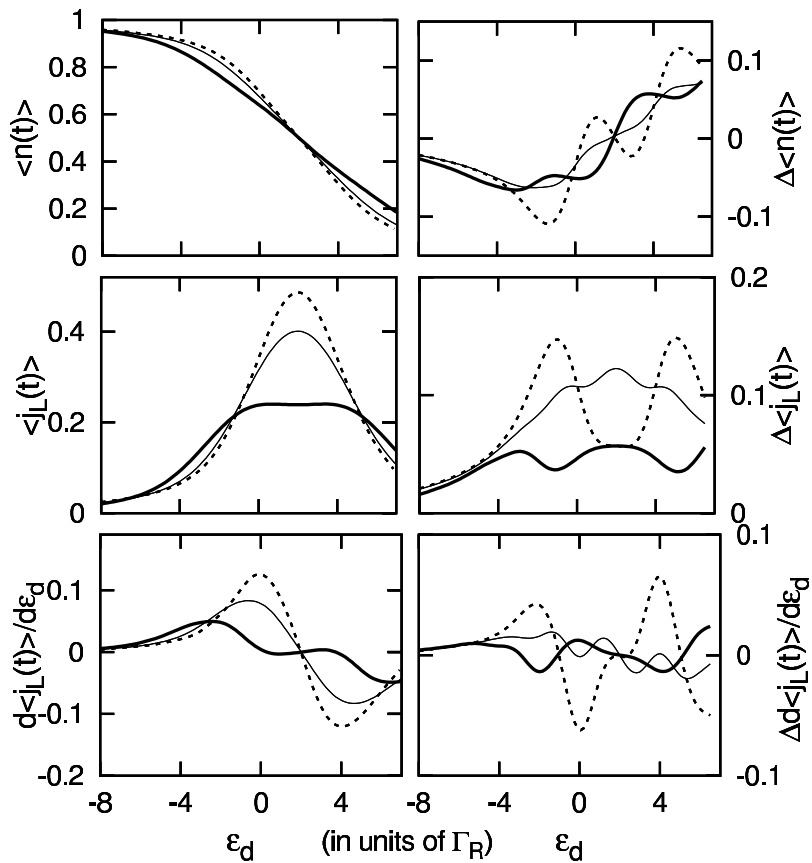}}
\end{center}
\caption{The time averaged QD charge $\langle n(t)\rangle$,
tunneling current $\langle j_L(t)\rangle$
and derivative $d\langle j_L(t)\rangle/d\epsilon_d$
(in arb. units) vs $\epsilon_d$ obtained for
2D-TB leads DOS - the left panels. In the right
 panels the corresponding differences  between the
 results showed in left panels and those
obtained within WBL ($\Delta \langle n(t)\rangle$,
$\Delta \langle j_L(t)\rangle$
and $\Delta \langle dj_L\rangle/ d\epsilon_d$) are given.
 The thick (thin) curves correspond to
$\Delta_L = 8$, $\Delta_d = 4$, $\Delta_R = 0$
($\Delta_L = 4$ ,$\Delta_d = 2$, $\Delta_R = 0)$
and $\omega = 2$, $\mu_R = 0$,
$\mu_L = 4$. The dashed curves correspond to the time-independent case. \cite{Lublin_qd3}}
\label{fig10}
\end{figure}

In Fig. \ref{fig10} in the left panels we show the time-averaged QD
 charge, tunneling current and the derivative of
the tunneling current with respect to the QD energy level
$\epsilon_d$. In the right panels the corresponding
differences ($\Delta \langle n(t)\rangle$, $\Delta \langle j_L(t)\rangle$
and $\Delta \langle dj_L\rangle/d\epsilon_d$) between the results
obtained for 2D-TB leads DOS and obtained
within wide-band limit (WBL) usually used in literature are given.
The dashed lines correspond to the static case ($\omega = 0$)
and the solid lines
correspond to the time-dependent transport.

%%%%%%%%%%%%%%%%%%%%%%%%%%%%%%%%%%%%%%%%%%%%%%%%%%%%%%%%%%%%%%%%%%%%%%%%%%%%
%%%%%%%%%%%%%%%%%%%%%%%%%%%%%%%%%%%%%%%%%%%%%%%%%

One can see, that, e.g. for $\epsilon_d = 1$, the tunneling
current calculated for 2D-TB leads DOS is
greater than that obtained within WBL up to a factor of
$\sim 0.30$ whilst for $\epsilon_d = 1$ up to $\sim 1.50$
for the static case and this difference
decreases with increasing amplitude $\Delta_L$
($\Delta_d = \Delta_L/2$, $\Delta_R = 0$). It is interesting that
the $\Delta \langle j_L(t) \rangle$ curves for different
$\Delta_L$ exhibit local minima or maxima at the same
$\epsilon_d$. For example, $\Delta \langle j_L(t) \rangle$
for the static case possesses a local minimum
for $\epsilon_d = 2$ (the middle point between the chemical
potentials $\mu_L$ and $\mu_R$) but for the time-dependent
case we observe local maxima, greater for
$\Delta_L = 4$ and lesser for greater amplitude  $\Delta_L = 8$.
The bottom panels show the derivative
$d\langle j_L(t)\rangle /d\epsilon_d$ vs $\epsilon_d$ and the
corresponding differences obtained for two considered
 models of the leads
DOS. As before, the greatest differences are observed
for the static case (dashed
lines) for QD energy levels $\epsilon_d$ equal to the chemical
potentials $\mu_L$ or $\mu_R$. For the time-dependent
transport these differences are much smaller
and hardly depend on amplitudes $\Delta_L$ and $\Delta_d$, contrary
to $\Delta \langle j_L(t)\rangle$ which are greater for smaller values of
$\Delta_L$ and $\Delta_d$.

%%%%%%%%%%%%%%%%%%%%%%%%%%%%%%%%%%%%%%%%%%%%%%%%%%%%%%%%%%%%%%%%%%%%%%%%%%%%%%%%%%%%%%%%%%%%%%%%%%%%%%%%%%%%%%%%%%%%%%%%%%%%%%%%%%
\begin{figure}[h]
\begin{center}
\resizebox{0.6\columnwidth}{!}{
\includegraphics{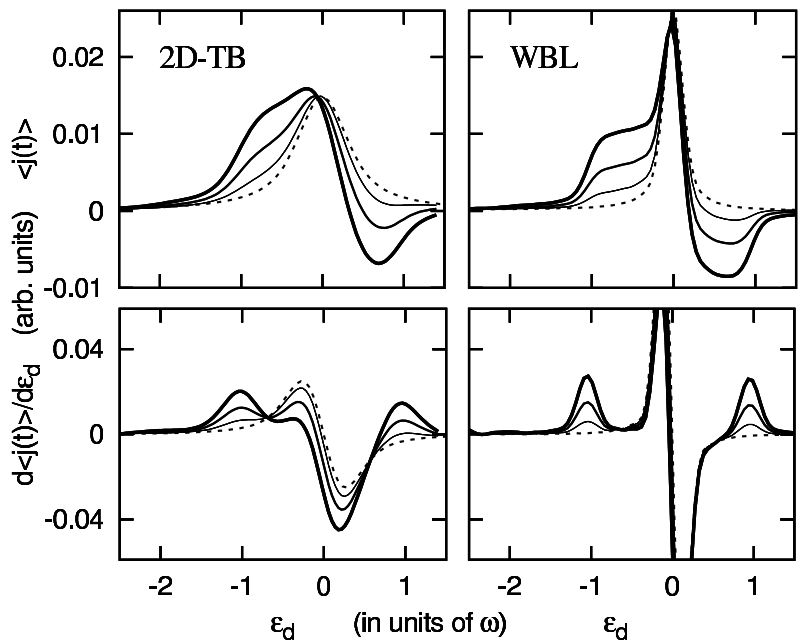}}
\end{center}
\caption{Left panels - the average current $\langle j_L(t)\rangle$ and
derivative $d\langle j_L(t)\rangle/d\epsilon_d$ vs $\epsilon_d$ for external field applied
only to the right lead in the case of the 2D-TB leads DOS. The solid curves correspond to $\Delta_R/\omega =
0.3$ (thin lines), 0.5 and 0.7 (thick lines). The dotted curve is the resonant peak without external field.
The parameters are $\Gamma_L = \Gamma_R = 0.1\omega$ and $\mu_L = - \mu_R = 0.05 \omega$.
For comparison, on the right panels the same as on the left but obtained within WBL. \cite{Lublin_qd3}}
\label{fig11}
\end{figure}
%%%%%%%%%%%%%%%%%%%%%%%%%%%%%%%%%%%%%%%%%%%%%%%%%%%%%%%%%%%%%%%%%%%%%%%%%%%%%%%%%

In the next step we consider the influence of the lead DOS structure
 on the photon-assisted tunneling
through a quantum dot. In experiment, in average current vs gate
voltage curve, a "shoulder" is observed on
the left side of the main resonant peak for the case of
a MW field applied on one lead only.
In Fig.\ref{fig11} we compare
 the structure of the main resonant peak in the average
  current $\langle j_L(t)\rangle$ and the
derivative $d\langle j_L\rangle/d\epsilon_d$ plotted vs $\epsilon_d$ obtained within
 WBL (right panels) and calculated for 2D-TB leads DOS (left panels) at half-filled bands.
 The remarkable differences are visible for negative values of $\epsilon_d$.
 The main resonant peaks are broader and lower and their height is almost independent on the amplitude
 $\Delta_R$. The photon-assisted tunneling for $\epsilon_d = \pm 1$ is clearly visible (lower panels of
 Fig.\ref{fig11}).

%%%%%%%%%%%%%%%%%%%%%%%%%%%%%%%%%%%%%%%%%%%%%%%%%%%%%%%%
\subsection{\label{subsec2}Electron transport through a QD in the presence of a direct
tunneling between leads}
%%%%%%%%%%%%%%%%%%%%%%%%%%%%%%%%%%%%%%%%%%%%%%%%%%%%%%%5

The progress of nanomaterials science has enabled
the experimental study of the phase coherence
of the charge carriers in many mesoscopic systems.
Especially, the Fano resonances in the
conductance were observed \cite{10}, which imply
that there are two paths for transfer of electrons
between a source and a drain. The recent
experimental and theoretical study with a low-temperature
scanning tunneling microscope (STM) of the single magnetic
atom deposited on a metallic surface showed also the asymmetric
Fano resonances in the tunneling spectra
\cite{10,11,12,13}. The STM measurements indicate that in tunneling
of electrons between STM tip and a surface with a single
impurity atom two different paths are present. The
electrons can tunnel between the tip and the adsorbate
state and directly between the tip and the metal surface.

In this subsection we address the issue of a QD
with a bridge channel between a source and a drain driven out
of equilibrium by means of a dc voltage bias and additional
time-dependent external fields.

We model the QD coupled to the left and right electron reservoirs with
the additional bridge tunneling channel between them
(represented by the coupling $V_{LR}$)
by the usually used Anderson-type Hamiltonian where $U=0$.

%%%%%%%%%%%%%%%%%%%%%%%%%%%%%%%%%%%%%%%%%%%%%%%%%%%%%%%%%%%%%%%%%%%%%%%%%%%%%%%%%%%%%%%%%%%%%%%%%%%%%%%%%%%%%%%%%%%%%%%%%%%%%%%%%%
\begin{figure}[]
\begin{center}
\resizebox{0.63\columnwidth}{!}{
\includegraphics[angle=-90]{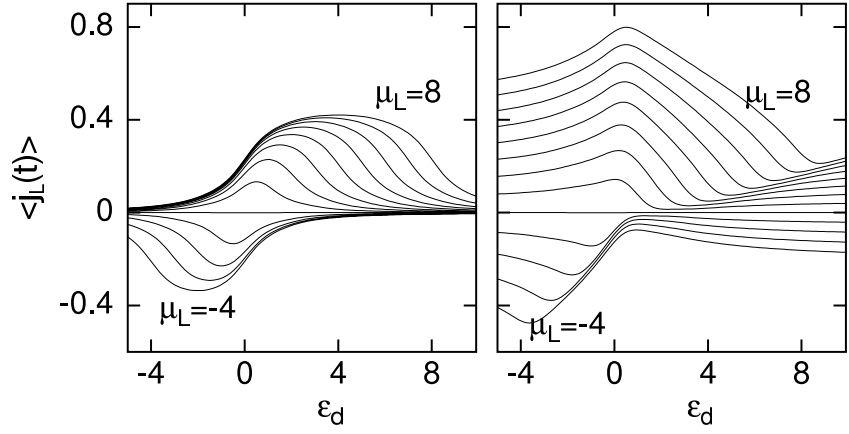}}
\end{center}
\caption{The average current $\langle j_L(t)\rangle$ against $\varepsilon_d$
for given values of $\mu_L$ (beginning from $\mu_L = -4$ up to $\mu_L = 8$). The left
and right panels correspond to $V_{RL} = 0$ and $V_{RL} = 10$, respectively,
$\mu_R = 0$, $V = 4$, $\Delta_L = 2$,
$\Delta_d = 1$, $\Delta_R = 0$, $\omega = 2$. \cite{Lublin_qd9}}
\label{fig12}
\end{figure}
%%%%%%%%%%%%%%%%%%%%%%%%%%%%%%%%%%%%%%%%%%%%%%%%%%%%%%%%%%%%%%%%%%%%%%%%%%%%%%%%%%%%%%%%%%%%%%%%%%%%%%%%%
%%%%%%%%%%%%%%%%%%%%%%%%%%%%%%%%%%%%%%%%%%%%%%%%%%%%%%%%%%%
\begin{figure}[t,h,b]
\begin{center}
\resizebox{0.63\columnwidth}{!}{
\includegraphics[angle=-90]{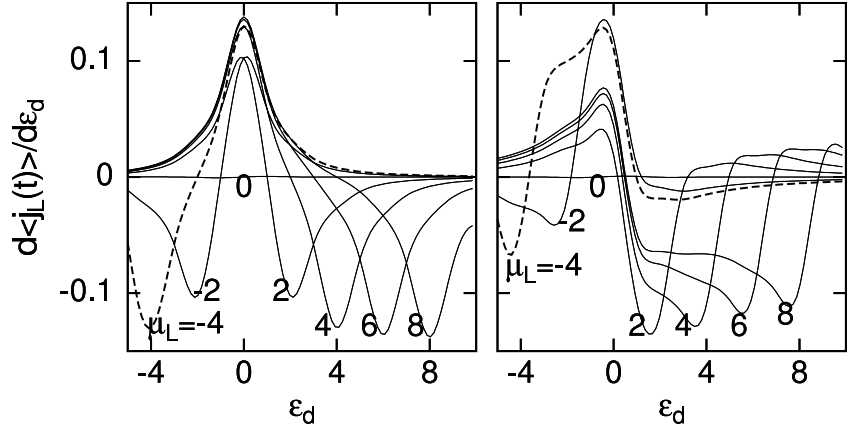}}
\end{center}
\caption{The derivatives of the average current against $\varepsilon_d$
with respect to the QD energy
level $\varepsilon_d$, $d\langle j_L(t)\rangle/d\varepsilon_d$,  for given values of $\mu_L$
(beginning from $\mu_L = -4$ up to $\mu_L = 8$). The broken curves correspond
to $\mu_L = -4$. The left  (right) panel
corresponds to $V_{RL} = 0$  ($V_{RL} = 10$) and the other
parameters as in Fig. \ref{fig12}. \cite{Lublin_qd9}.}
\label{fig13}
\end{figure}
%%%%%%%%%%%%%%%%%%%%%%%%%%%%%%
%%%%%%%%%%%%%%%%%%%%%%%%%%%%%%%%%%%%%%%%%%%%%%%%%%%%%%%%%%%%%%%55
\begin{figure}[h]
\begin{center}
\resizebox{0.63\columnwidth}{!}{
\includegraphics[angle=-90]{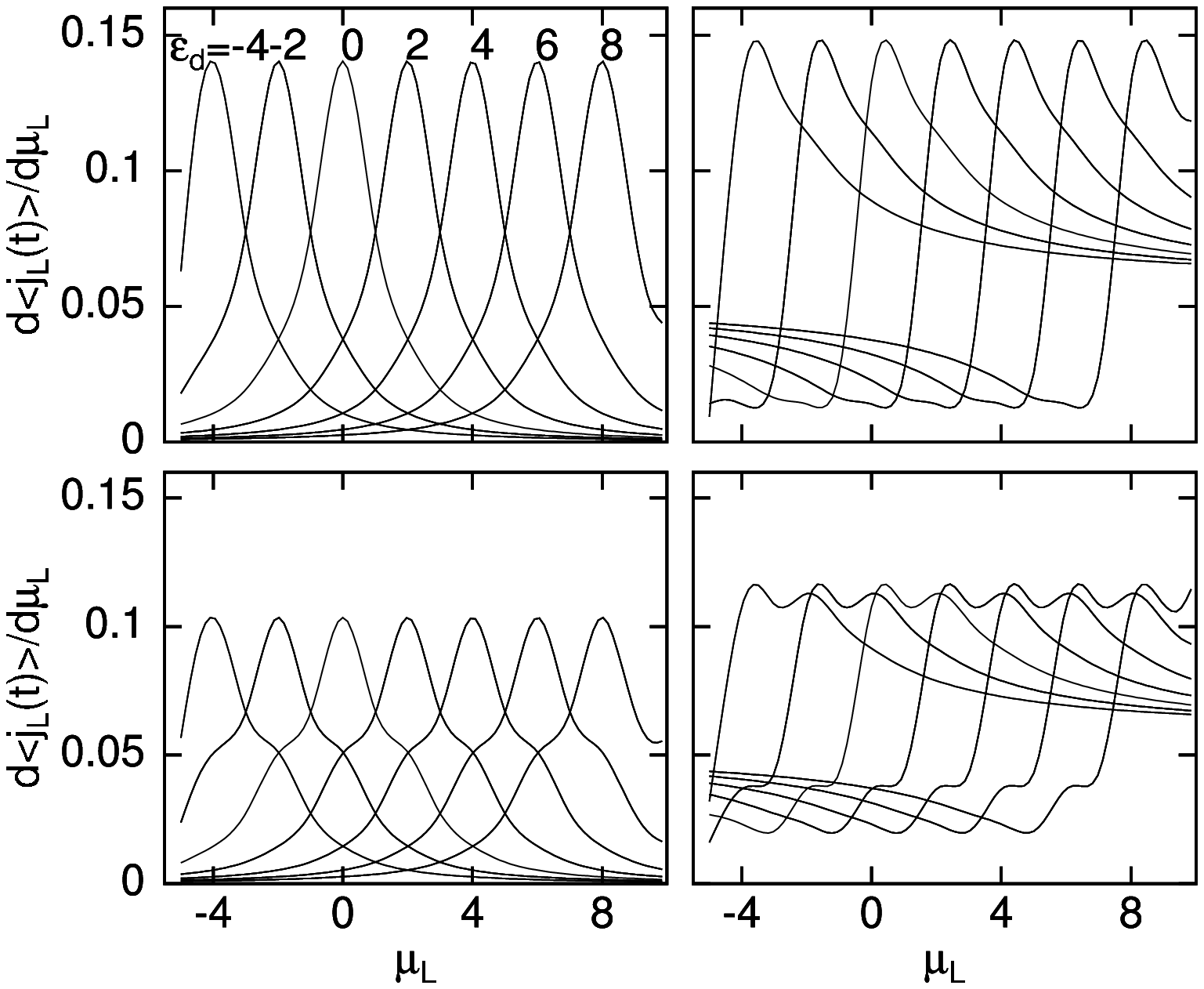}}
\end{center}
\caption{The derivatives of the average current against $\mu_L$
with respect to $\mu_L$, $d\langle j_L(t)\rangle/d\mu_L$, for
given values of $\varepsilon_d$ (beginning from $\varepsilon_d =
-4$ up to $\varepsilon_d = 8$). The left (right) panels correspond
to $V_{RL} = 0$ ($V_{RL} = 10$) and upper (lower) panels
correspond to $\Delta_L = 2$ ($\Delta_L = 4$). The other
parameters as in Fig. \ref{fig12}. \cite{Lublin_qd9}}
\label{fig14}
\end{figure}
%%%%%%%%%%%%%%%%%%%%%%%%%%%%%%%%%%%%%%%%%%%%%%%%%%%%%%%%%%%%%%%%%%%%%%%%%%%%%%%%%%%%%%%%%%%%%%%%%%%%%%%%%

Here we show the numerical results of the time-averaged
current $\langle j_L(t)\rangle$ and its derivatives with
respect to the QD energy level position and the chemical
potential $\mu_L$ (or equivalently, with respect to the
gate and source-drain voltages) for different sets
of parameters which characterize our system.

At first, we consider
the dependence of $\langle j_L(t)\rangle$ vs. $\varepsilon_d$ for given values of the
left lead chemical potential $\mu_L$. In Fig. \ref{fig12} we present
such curves for different values of $\mu_L$  obtained for $\mu_L = 8$.
The left (right) panel corresponds
to $V_{RL} = 0$ ($V_{RL} = 10$). In the case of vanishing over-dot tunneling
channel (the left panel) the current has a simple structure --
a single peak localized in the middle between $\mu_L$ and $\mu_R$
for smaller values of $\mu_L$. The width of this peak increases with
increasing $|\mu_L|$ and for greater values of $|\mu_L|$ the current
is almost independent of $\varepsilon_d$ localized inside the energy region
between $\mu_R$ and $\mu_L$.
For non-vanishing  over-dot tunneling (Fig. \ref{fig12}, the right panel)
the curves $\langle j_L(t)\rangle$ become asymmetric.
With increasing source-drain bias, the current
possesses greater values
in comparison with the $V_{RL} = 0$ case due to the
direct tunneling between both leads. Note, however, that due to the
interference effects the resulting
$\langle j_L(t)\rangle$ curves are asymmetric. The interference effects
are most visible for $\varepsilon_d$ lying approximately in the region ($\mu_R , \mu_L$).

In Fig. \ref{fig13} we show the derivatives of the average current
vs. the QD energy level $\varepsilon_d$ obtained for some values of $\mu_L$.
There are the results of the differentiation of curves shown in Fig. \ref{fig12}.
Again, the most visible differences between the results obtained
for $V_{RL} = 0$ and $V_{RL} \neq 0$ are present for the QD energy level
$\varepsilon_d$ localized approximately between chemical
potentials of both leads (compare, for example, the curves calculated
for $\mu_L = 8$).

Fig. \ref{fig14} presents the comparison of the
$d\langle j_L(t)\rangle/d\mu_L$ vs.
$\mu_L$ curves calculated for vanishing $V_{RL}$ (left panels)
and for $V_{RL} = 10$ (right panels) for two different values
of the amplitudes $\Delta_L$ ($\Delta_d = \Delta_L/2$, $\Delta_R = 0$).
At the vanishing value of $V_{RL}$, the
shape of the curves is symmetrical in relation
to the values $\mu_L = \varepsilon_d$ although for greater $\Delta_L$
some shoulders appear on both sides of the
corresponding peaks in the distance $\sim \Delta_L/2$ from the curve
centers. For non-vanishing $V_{RL}$, the corresponding
curves are approximately asymmetric and for
large values of $\mu_L$ they tend to constant, non-zero values
corresponding to linear increasing of the current at
large $\mu_L$. It is interesting that with the increasing
amplitudes $\Delta_L$
and $\Delta_d$ very clear structures appear on
both sides of the corresponding curves. Note, that
all these curves can be obtained, for example, from
the one calculated for $\varepsilon_d = 0$ and moved along
the $\mu_L$-axis by the corresponding value (equal to $\varepsilon_d$).

%%%%%%%%%%%%%%%%%%%%%%%%%%%%%%%%%%%%%%%%%%%%%%%%%%%%%%%%%%%%
\begin{figure}[h]
\begin{center}
\resizebox{0.63\columnwidth}{!}{
\includegraphics{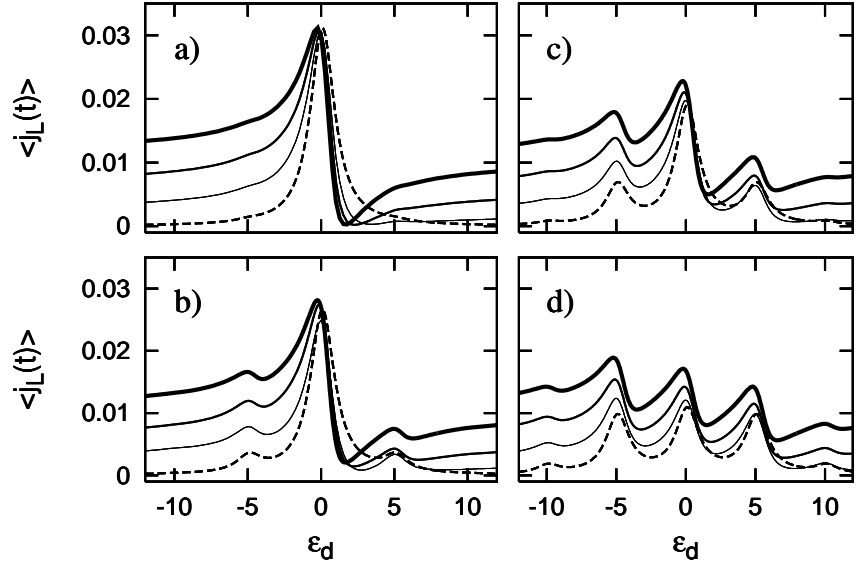}}
\end{center}
\caption{The average current $\langle j_L(t)\rangle$ against $\varepsilon_d$ for
the oscillating QD energy level at $V_{RL}$ = 0, 4, 7, 10 -- broken, thin, thick and
very thick curves, respectively. The panels a, b, c and d correspond to
$\Delta_d = 1, 3, 5$ and 7, respectively. $\omega = 5$, $\Gamma = 1$, $V = 4$,
$\mu_L = 0.2$, $\Delta_L = \Delta_R = 0$. \cite{Lublin_qd9}}
\label{fig15}
\end{figure}
%%%%%%%%%%%%%%%%%%%%%%%%%%%%%%%%%%%%%%%%%%%%%%%%%%%%%%%%%%%%%%%%%%%%%%%%%%%%%%%%%%%%%%%%%%%%%%%%%%%%%%%%%
%%%%%%%%%%%%%%%%%%%%%%%%%%%%%%%%%%%%%%%%%%%%%%%%%%%%%%%%%%%%%%%%%%
\begin{figure}[h]
\begin{center}
\resizebox{0.45\columnwidth}{!}{
\includegraphics{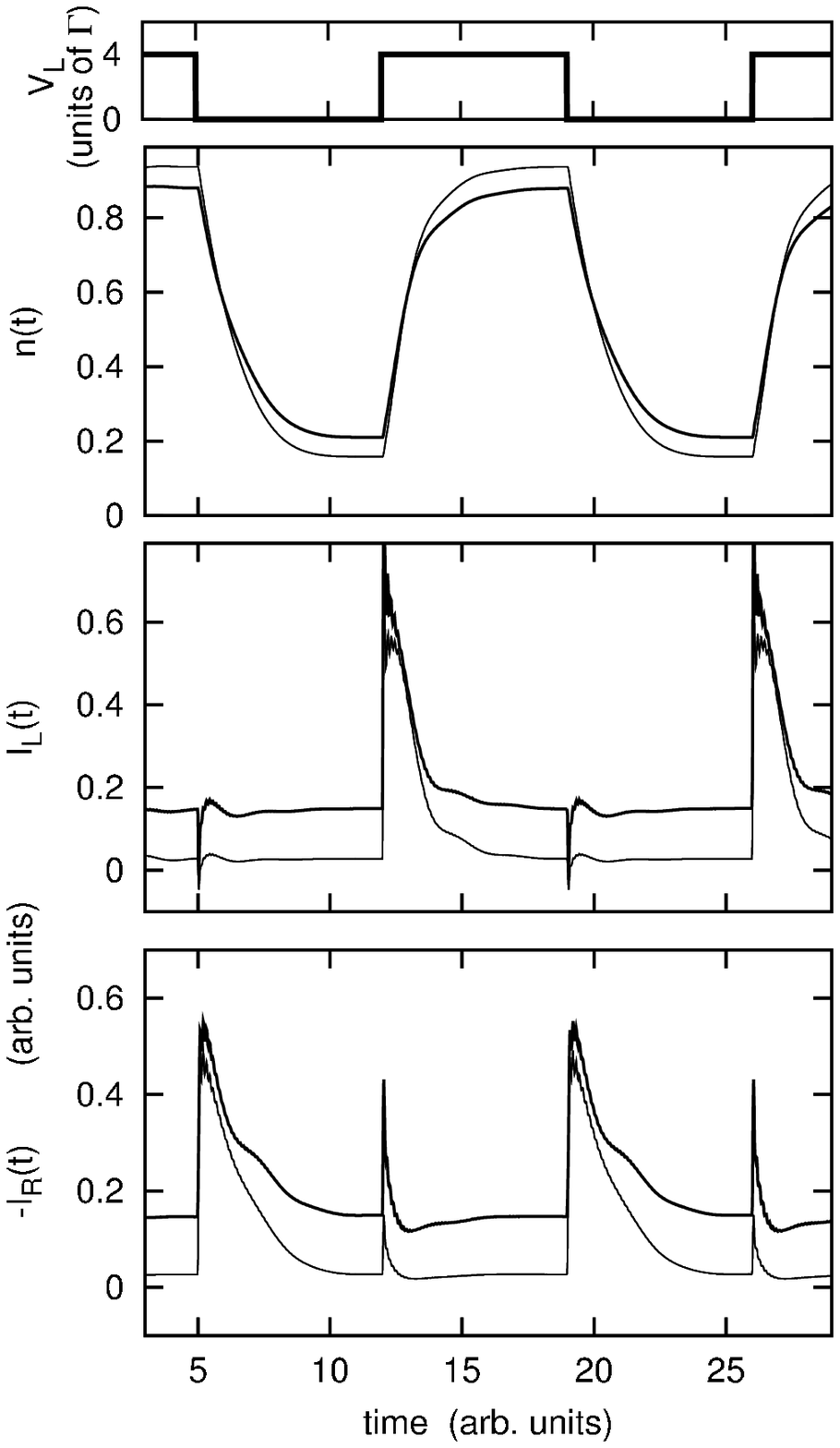}}
\end{center}
\caption{The time-dependence of $n(t)$, $I_L(t)$ and $I_R(t)$ for
the case when the QD-lead barriers are changing in time as shown
for $V_L(t)$ in the upper panel ($V_R(t)$ is out of phase with a
phase difference of $\pi$) for two values of the over-dot bridge
tunneling channel strength corresponding to $V_{RL} = 4$ (thin
lines) and $V_{RL} = 10$ (thick lines). $\varepsilon_d = 1$, $V =
4$, $\mu_L = 4$, $\mu_R = 0$. \cite{Lublin_qd1}} \label{fig16}
\end{figure}
%%%%%%%%%%%%%%%%%%%%%%%%%%%%%%%%%%%%%%%%%%%%%%%%%%%%%%%%%%%%

Fig. \ref{fig15} presents the average current
$\langle j_L(t)\rangle$ obtained
for the case in which  only the QD energy
level $\varepsilon_d$ oscillates with some frequency ($\omega > \Gamma$ )
and for the small source-drain
voltage $\mu_L-\mu_R = 0.2$. For vanishing $V_{RL}$
 (broken curves) we observe for small
amplitude $\Delta_d$ only a central resonant peak (Fig. \ref{fig15}a).
With increasing $\Delta_d$, the subsequent
peaks appear and the distance between them
and the central peak is an integer multiple of the frequency $\omega$ (side bands).
The location of peaks is independent of the
amplitude $\Delta_d$ but their relative intensity values change and
with increasing $\Delta_d$ the height of the central peak
is reduced. If we take the
additional over-dot tunneling channel into consideration,
especially for small $\Delta_d$, the asymmetric shape of the
current curve is observed and this asymmetry increases with the
increasing strength of the over-dot coupling
between both leads. With the increasing amplitude
$\Delta_d$ this asymmetry is reduced largely due to the
extra, photon-assisted tunneling peaks whose strength
increases with the increasing $\Delta_d$.

As a last problem we consider the QD with the periodic rectangular-pulse
external field applied to each QD-lead
barrier. We assume that the influence of these fields on the
system under consideration is equivalent to the following time-dependence of
the matrix elements $V_L$ and $V_R$ corresponding to the coupling of the QD
with both leads:
 \begin{eqnarray}
 V_L(t) = 4,  V_R = 0 \,\, for \,\, 0 \leq t < T/2 \nonumber\\
 V_L(t) = 0,  V_R = 4  \,\, for \,\, T/2 \leq t < T .\nonumber
\end{eqnarray}

The bridge tunneling channel is present all the time and the
calculations were performed for $V_{RL} = 4$ and $V_{RL} = 10$.
In the upper part of Fig. \ref{fig16} we show the time-dependence of $V_L(t)$.
During the periodic coupling of the QD to the left or right lead the
QD charge oscillates and the amplitude of these
oscillations decreases with the increasing coupling $V_{RL}$ between both
leads. Note, that the greater $V_{RL}$ enhances the equilibrium value
of the QD charge at $V_L = 0$ and $V_R \neq 0$ (in the first
half-period), but reduces it for $V_L \neq 0$ and $V_R = 0$ (i.e. in
the second half-period). In two lower parts of Fig. \ref{fig16} we show the
current leaving the right and left leads for two values of the coupling
$V_{RL}$. The effect
of switching-on and -out of the QD-leads coupling on the currents
is markedly visible. The time after which the currents achieve their
steady values hardly depends on the coupling $V_{RL}$, although the width
of the corresponding current peaks is smaller for lower values of
$V_{RL}$.

\newpage
\section
{Concluding remarks: Is all this related to quantum information
processing?}

The problems discussed here are of importance for
small structures working at low temperatures. In such cases
the presence of  even single electron on the structure prevents
other electrons to tunnel onto it. This is because small capacitance
of these systems results in large charging energies which have been
parametrised here by the Coulomb on dot repulsion.

As the ultimate goal of the miniaturisation is to build
a quantum computer it maybe of interest to ask the question whether
the systems considered will ever be used for this purpose.
The answer is affirmative. It has been proposed to
use  charge \cite{Brum}  or spin \cite{loss}
on a quantum dot to build a qubit -- an elementary quantum unit of
information. The subsequent work \cite{Mun} has shown
 that the decoherence of  a system
 consisting of two lateral quantum dots due to cotunneling does depend on
the difference between chemical potentials of the two leads
i.e. on the voltage. The future work will show whether
the control of decoherence by the external bias will
also be effective in more complex geometries.

\newpage
{\bf Acknowledgements:} This work has been
partially supported by the grant no. PBZ-MIN-008/P03/2003 .

%%%%%%%%%%%%%%%%%%%%%%%%%%%%%%%%%%%%%%%%%%%%%%%%%%%%%%%%%%%%%%%%%%%%%%%%%%%%%%%

\end{document}